\documentstyle[11pt,psfig,twoside]{article}

\begin{document}

\newcommand{\dd}{\,{\rm d}}
\newcommand{\ie}{{\it i.e.},\,}
\newcommand{\etal}{{\it et al.\ }}
\newcommand{\eg}{{\it e.g.},\,}
\newcommand{\cf}{{\it cf.\ }}
\newcommand{\vs}{{\it vs.\ }}
\newcommand{\zdot}{\makebox[0pt][l]{.}}
\newcommand{\up}[1]{\ifmmode^{\rm #1}\else$^{\rm #1}$\fi}
\newcommand{\dn}[1]{\ifmmode_{\rm #1}\else$_{\rm #1}$\fi}
\newcommand{\upd}{\up{d}}
\newcommand{\uph}{\up{h}}
\newcommand{\upm}{\up{m}}
\newcommand{\ups}{\up{s}}
\newcommand{\arcd}{\ifmmode^{\circ}\else$^{\circ}$\fi}
\newcommand{\arcm}{\ifmmode{'}\else$'$\fi}
\newcommand{\arcs}{\ifmmode{''}\else$''$\fi}
\newcommand{\MS}{{\rm M}\ifmmode_{\odot}\else$_{\odot}$\fi}
\newcommand{\RS}{{\rm R}\ifmmode_{\odot}\else$_{\odot}$\fi}
\newcommand{\LS}{{\rm L}\ifmmode_{\odot}\else$_{\odot}$\fi}

\newcommand{\Abstract}[2]{{\footnotesize\begin{center}ABSTRACT\end{center}
\vspace{1mm}\par#1\par
\noindent
{~}{\it #2}}}

\newcommand{\TabCap}[2]{\begin{center}\parbox[t]{#1}{\begin{center}
  \small {\spaceskip 2pt plus 1pt minus 1pt T a b l e}
  \refstepcounter{table}\thetable \\[2mm]
  \footnotesize #2 \end{center}}\end{center}}

\newcommand{\TableSep}[2]{\begin{table}[p]\vspace{#1}
\TabCap{#2}\end{table}}

\newcommand{\FigCap}[1]{\footnotesize\par\noindent Fig.\  %
  \refstepcounter{figure}\thefigure. #1\par}

\newcommand{\TableFont}{\footnotesize}
\newcommand{\TableFontIt}{\ttit}
\newcommand{\SetTableFont}[1]{\renewcommand{\TableFont}{#1}}

\newcommand{\MakeTable}[4]{\begin{table}[htb]\TabCap{#2}{#3}
  \begin{center} \TableFont \begin{tabular}{#1} #4 
  \end{tabular}\end{center}\end{table}}

\newcommand{\MakeTableSep}[4]{\begin{table}[p]\TabCap{#2}{#3}
  \begin{center} \TableFont \begin{tabular}{#1} #4 
  \end{tabular}\end{center}\end{table}}

\newenvironment{references}%
{
\footnotesize \frenchspacing
\renewcommand{\thesection}{}
\renewcommand{\in}{{\rm in }}
\renewcommand{\AA}{Astron.\ Astrophys.}
\newcommand{\AAS}{Astron.~Astrophys.~Suppl.~Ser.}
\newcommand{\ApJ}{Astrophys.\ J.}
\newcommand{\ApJS}{Astrophys.\ J.~Suppl.~Ser.}
\newcommand{\ApJL}{Astrophys.\ J.~Letters}
\newcommand{\AJ}{Astron.\ J.}
\newcommand{\IBVS}{IBVS}
\newcommand{\PASP}{P.A.S.P.}
\newcommand{\Acta}{Acta Astron.}
\newcommand{\MNRAS}{MNRAS}
\renewcommand{\and}{{\rm and }}
\section{{\rm REFERENCES}}
\sloppy \hyphenpenalty10000
\begin{list}{}{\leftmargin1cm\listparindent-1cm
\itemindent\listparindent\parsep0pt\itemsep0pt}}%
{\end{list}\vspace{2mm}}

\def\TYLDA{~}
\newlength{\DW}
\settowidth{\DW}{0}
\newcommand{\dw}{\hspace{\DW}}

\newcommand{\refitem}[5]{\item[]{#1} #2%
\def\REFARG{#3}\ifx\REFARG\TYLDA\else, {\it#3}\fi
\def\REFARG{#4}\ifx\REFARG\TYLDA\else, {\bf#4}\fi
\def\REFARG{#5}\ifx\REFARG\TYLDA\else, {#5}\fi.}

\newcommand{\Section}[1]{\section{#1}}
\newcommand{\Subsection}[1]{\subsection{#1}}
\newcommand{\Acknow}[1]{\par\vspace{5mm}{\bf Acknowledgements.} #1}
\pagestyle{myheadings}

\def\thefootnote{\fnsymbol{footnote}}
\begin{center}
{\large\bf The Optical Gravitational Lensing Experiment.\\
Red Clump Stars as a Distance Indicator\footnote
{Based on  observations obtained with the 1.3~m Warsaw telescope at the
Las Campanas  Observatory of the Carnegie Institution of Washington.}}

\vskip0.8cm
{\bf
A.~~U~d~a~l~s~k~i}
\vskip3mm
{Warsaw University Observatory, Al.~Ujazdowskie~4, 00-478~Warszawa, Poland\\
e-mail: udalski@astrouw.edu.pl}
\end{center}

\Abstract{
We present relation of the mean {\it I}-band brightness of red clump
stars on metallicity. Red clump stars were proposed to be a very
attractive standard candle for distance determination. The calibration
is based on 284 nearby red giant stars whose high quality spectra
allowed to determine accurate individual metal abundances. High quality
parallaxes ($\sigma_{\pi}/\pi < 10\%$) and photometry of these very
bright stars  come from Hipparcos measurements. Metallicity of the
sample covers a large range: $-0.6<{\rm [Fe/H]}<+0.2$~dex. We find  a
weak dependence of the mean {\it I}-band brightness on metallicity
($\approx0.13$~mag/dex).

What is more important, the range of metallicity of the Hipparcos sample
partially overlaps with metallicity of field giants in the LMC, thus
making it possible to determine the distance to the LMC by almost direct
comparison of brightness of the local Hipparcos red clump giants with
that of LMC stars. Photometry of field red clump giants in nine low
extinction fields of the LMC halo collected during the OGLE-II
microlensing survey compared with the Hipparcos red clump stars data
yields the distance modulus to the LMC: $(m-M)_{\rm
LMC}=18.24\pm0.08$~mag.
}{~}

\Section{Introduction}

Paczy{\'n}ski and Stanek (1998, hereafter PS) noticed that the {\it
I}-band absolute brightness, $M_I$, of red clump (hereafter RC) giants,
the intermediate age (2--10~Gyr) helium core burning stars, has
intrinsically small dispersion and can be used as a "standard candle"
for distance determination. The obvious advantage of this method is 
large number of these stars in stellar populations  allowing
determination of the mean brightness with statistically unprecedented
accuracy. Moreover, RC giants are also very numerous in the solar
neighborhood. Therefore their brightness could be precisely calibrated
with hundreds of stars for which the Hipparcos satellite measured
parallaxes with accuracy better than 10\% (Perryman \etal 1997). It
should be noted that RC stars are the only standard candle which can be
calibrated with direct, trigonometric parallax measurements. Similar
quality parallaxes do not exist for Cepheids or RR~Lyr stars (\cf Fig.~2
of Horner \etal 1999).

As in the case of any other stellar standard candle, the  brightness of
RC stars might, however,  be affected by population effects: different
chemical composition or age of the stellar system being studied, as
compared to the local Hipparcos giants. PS argued that both dependences
of $\langle M_I\rangle$  on age and metallicity are negligible. On the
other hand,  Girardi \etal (1998) claimed much stronger dependences
based on theoretical modeling. According to their results $\langle
M_I\rangle$ of RC stars could be different by as much as 0.5~mag in
different environments. However, one should be aware that  results of
modeling  are quite sensitive to the input physics and results from
different types of evolutionary codes are not consistent each other
(Dominguez \etal 1999, Castellani \etal 1999). While models seem to
reasonably reproduce qualitative  properties of RC stars they cannot
provide accuracy of hundredths of magnitude  required for precise
distance determination.

The necessity of good, preferably empirical, calibration of population
effects on RC brightness became very urgent, in particular when the
distance modulus to the LMC was determined (Udalski \etal 1998, Stanek,
Zaritsky and Harris 1998), supporting the "short" distance scale to that
galaxy ($\approx 0.4$~mag smaller than the "long" value of $(m-M)_{\rm
LMC}=18.50$~mag).  The distance to the LMC is one of the most important
distances of the modern astrophysics because extragalactic distance
scale is tied to it. Udalski (1998a) presented an empirical calibration
of $\langle M_I\rangle$ of RC stars on metallicity suggesting only a
weak dependence ($0.09\pm0.03$~mag/dex). The dependence of $\langle
M_I\rangle$ of RC stars on age was studied by Udalski (1998b).
Observations of RC stars in several star clusters of different age
located in low extinction areas of the Magellanic Clouds showed that
their $\langle M_I\rangle$ in these clusters is independent of age (for
stars 2--10~Gyr old) within observational uncertainties of a few
hundredths of magnitude. Recently, Sarajedini (1999) presented analysis
of a few Galactic open clusters suggesting fainter RC in older
($>$5~Gyr) clusters.

In this paper we present empirical arguments which additionally support
usefulness of the "RC stars" method -- the relation of $\langle
M_I\rangle$ \vs metallicity based on the most precise data necessary to
solve the problem: high resolution and S/N spectra of nearby red giants,
for which accurate  parallaxes and photometry were measured by
Hipparcos. Large range of metallicity of nearby RC giants  partially
overlapping with metallicity range of field RC giants  in the LMC
enables us to compare brightness of these stars and determine the RC
distance modulus to the LMC  largely free from population uncertainties.

\Section{Observational Data}

The sample of red giant stars from the solar neighborhood comes from the
Hipparcos catalog and consists of objects with high accuracy
trigonometric parallaxes ($\sigma_{\pi}/\pi < 10\%$). About 75\% stars
from this sample are the same objects which were used by PS, \ie stars
with {\it I}-band photometry. To enlarge that data set we have also
included stars which {\it I}-band magnitude was obtained from $B-V$
color {\it via} very well defined correlation between $B-V$ and $V-I$
colors (\cf Fig.~8 Paczy{\'n}ski \etal 1999), \ie stars marked as type
'H' in the Hipparcos catalog. While accuracy of the {\it I} magnitude of
these stars is somewhat worse, it is still acceptable taking into
account usually larger uncertainty from parallax error.

\begin{figure}[htb]
\hglue1.5cm\psfig{figure=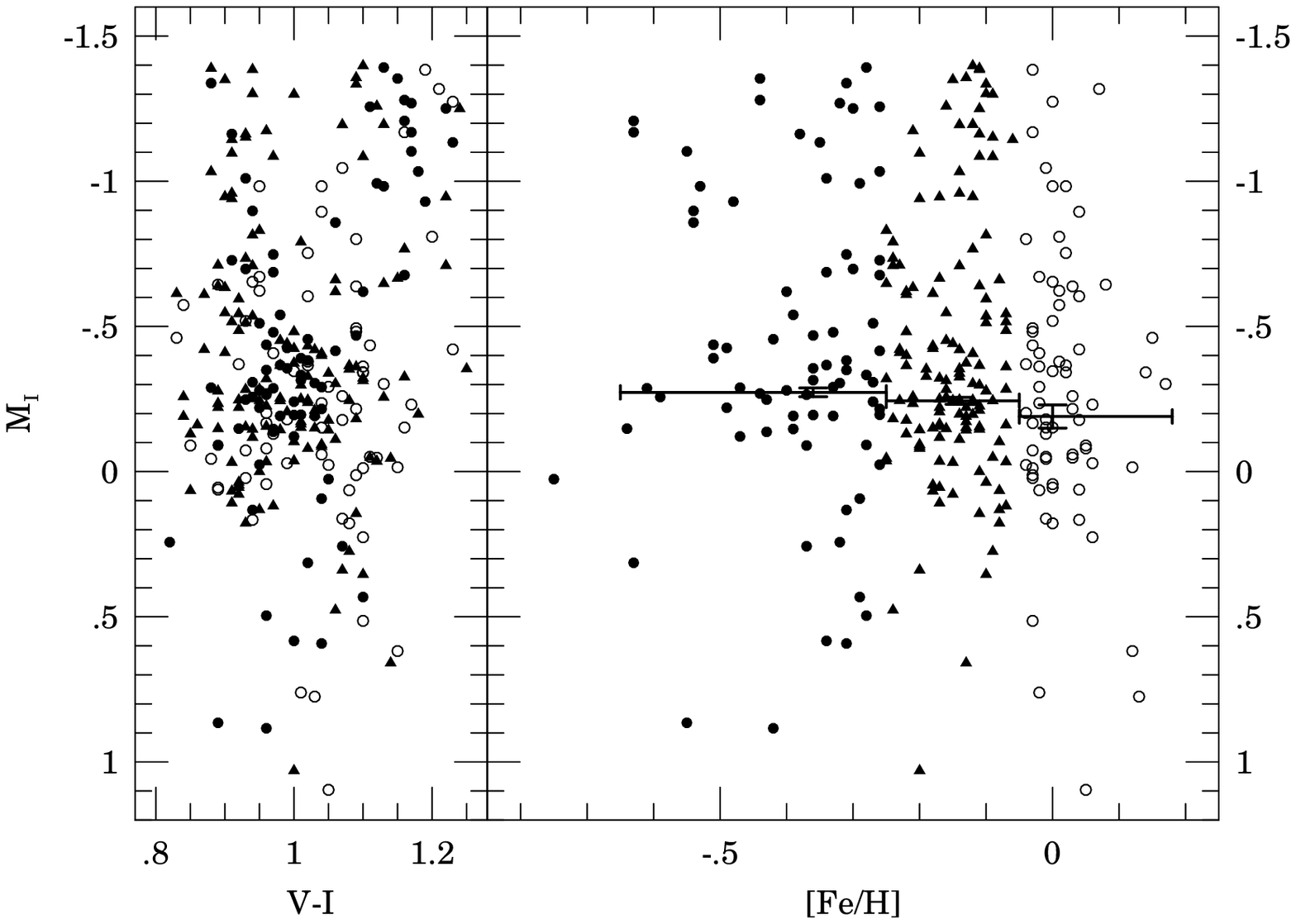,bbllx=45pt,bblly=60pt,bburx=530pt,bbury=420pt,width=10cm,clip=}
\vspace*{3pt}
\FigCap{$M_I$ of 284 nearby red giants with precise photometry and spectroscopy
plotted as a function of $V-I$ color (left panel) and metallicity [Fe/H]
(right panel). Stars of low, medium and high metallicity are marked by
filled circles, filled triangles and open circles, respectively.
}
\end{figure}

For further analysis only the stars with [Fe/H] abundance determination
were selected. We used results of the spectroscopic survey of McWilliam
(1990), containing [Fe/H] determinations for 671 G and K giants based on
high resolution ($R\approx40000$) and S/N ($\approx100$) spectra. This
is the most comprehensive and homogeneous data set available.  Typical
accuracy of [Fe/H] determination is about 0.1~dex. 284 objects from our
photometric sample were cross-identified with objects in the
spectroscopic survey list of McWilliam (1990). Left panel of Fig.~1
presents the color-magnitude diagram (CMD) of this sample. Comparison
with Fig.~2 of PS indicates that full photometric sample and our 284
object sample used for further analysis are distributed identically in
the CMD, so we did not introduce any significant bias when limiting
Hipparcos stars to objects with spectroscopic data. The mean distance
from the Sun of our final sample of 284 stars is $<d>=66$~pc. It is
worth noting that due to accurate Hipparcos parallaxes the possible
Lutz-Kelker bias of the absolute magnitude of our sample is negligible,
as shown by Girardi \etal (1998).

The photometric data of the LMC fields were obtained during the OGLE-II
microlensing survey (Udalski, Kubiak and Szyma{\'n}ski 1997).
Observations were collected with the 1.3-m Warsaw telescope at the Las
Campanas Observatory, Chile  which is operated by the Carnegie
Institution of Washington on eight photometric nights  between October
30, 1999 and November~12, 1999. Single chip CCD camera with $2048\times
2048$ pixel SITe thin chip was used giving a scale of 0.417
arcsec/pixel. $3-5$ frames in the {\it V} and {\it I}-bands were
collected for each field with exposure time of 300 sec for both bands.
Photometry was derived using the standard OGLE pipeline.  On each night
several standard stars from the Landolt (1992) fields were observed for
transformation of the instrumental magnitudes to the standard system.
The error of zero points of photometry should not exceed 0.02~mag.

\Section{Discussion}

\Subsection{Hipparcos Red Clump Stars}

CMD of the analyzed sample of red giants presented in the left panel of
Fig.~1 indicates that the majority of objects are RC stars that form a
very  compact structure located in the range: $0.0<M_I<-0.5$ and
$0.9<V-I<1.1$. They are, however, contaminated by red giant branch stars
and a blue vertical structure, going from $M_I\approx -1.4$~mag down to
$M_I\approx 0.1$~mag, consisting of younger (more massive) red giants.

$M_I$ of stars is plotted against metallicity [Fe/H] in the right panel
of Fig.~1. The distribution of stars is non-uniform with most objects
within metallicity range of $-0.3<{\rm [Fe/H]}<0.0$~dex. To investigate 
the relation of $M_I$ of RC stars on metallicity we divided our sample
into three subsamples: high metallicity, ${\rm [Fe/H]}>-0.05$~dex,
medium metallicity, $-0.25<{\rm [Fe/H]}<-0.05$~dex, and low metallicity
stars, ${\rm [Fe/H]}<-0.25$~dex. Such a division ensures more or less
uniform distribution of objects within each bin with  the mean, median
and mode values equal to ($0.02,0.00,0.00$), ($-0.15,-0.14,-0.14$) and
($-0.39,-0.36,-0.33$)~dex for the high, medium and low metallicity bins,
respectively. We determined $\langle M_I\rangle$  in all bins in similar
manner as described in Udalski \etal (1998). Fig.~2 presents the
histograms of distribution of $M_I$ for each subsample with fitted
Gaussian function representing RC stars superimposed on parabola
representing background stars.

\begin{figure}[htb]
\hglue 2.0cm\psfig{figure=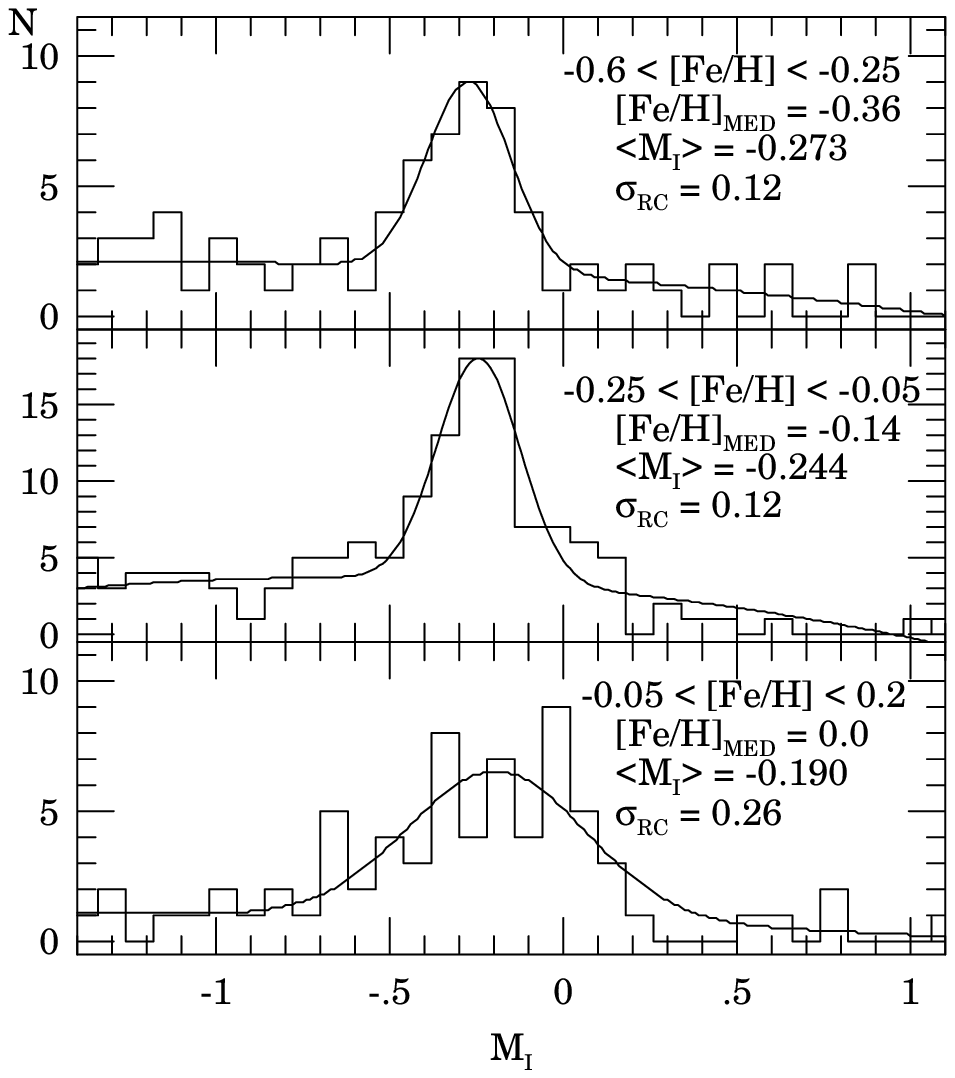,bbllx=75pt,bblly=125pt,bburx=355pt,bbury=435pt,width=8.5cm,clip=}
\vspace*{0pt}
\FigCap{
Histograms of distribution of $M_I$ of the low (top panel), medium
(middle panel) and high metallicity sample (bottom panel) of local RC
stars. Bins are 0.07~mag wide.
}
\end{figure}

It is evident from Fig.~2 that the majority of stars in the low and
medium metallicity bins are typical intermediate age RC objects. Small
dispersion of their $M_I$ ($\sigma_{\rm RC}=0.12$~mag) indicates that
these stars can indeed be a good standard candle. On the other hand the
high metallicity bin  has the RC poorly defined ($\sigma_{\rm
RC}=0.26$~mag). Closer examination of Fig.~1 indicates that most of
stars from this bin (open circles) are red branch giants and objects
located on the vertical sequence of younger giants in the RC evolution
phase with very few stars belonging to the intermediate age RC. 

$\langle M_I\rangle$ of RC stars in the low, medium and high metallicity
bins are equal to $-0.273\pm0.015, -0.244\pm0.012$ and
$-0.190\pm0.041$~mag, respectively (statistical error). The horizontal
bars in the right panel of Fig.~1 show the range of each bin and its
$\langle M_I\rangle$. The vertical bars are shown at the median
metallicity of each bin and represent statistical uncertainty of
$\langle M_I\rangle$. It is clear from Figs.~1 and 2 that the absolute
brightness of RC stars increases with lower metallicity. If we consider
all three bins then the slope of the $M_I$ \vs [Fe/H] relation is about
0.2~mag/dex. One should, however, remember that the high metallicity bin
is poorly defined due to small number of metal rich RC stars of
intermediate age in the solar neighborhood. Moreover, metallicity of the
most metal rich stars from McWilliam's sample is very likely
underestimated (McWilliam 1997) making this bin additionally uncertain.
Larger mean metallicity of this bin would lead to smaller slope of the
$M_I$ \vs [Fe/H] relation. Indeed, if we limit ourselves only to low and
medium metallicity bins where the intermediate age RC stars dominate the
linear relation becomes:
$$ M_I=(0.13\pm0.07)\cdot({\rm [Fe/H]}+0.25)- (0.26\pm0.02) \eqno{(1)} $$

Eq.~(1) indicates that the dependence of $M_I$ on metallicity is rather
weak. The relation is in good  agreement with the previous empirical
determination by Udalski (1998a) based on comparison of RC stars with
RR~Lyr stars but this time it is based on precise measurements of
numerous sample of individual stars, thus it is more reliable. It should
be also stressed that the result is weakly sensitive to systematic
errors, as those are very unlikely in the Hipparcos absolute photometry
of so bright and nearby stars, and due to weak dependence on metallicity
even large systematic metallicity error (which is also unlikely, Taylor
1999) would only lead to a magnitude shift of the order of a few
hundredths of magnitude.

\MakeTable{lcccrccc}{12.5cm}{Photometry of field red clump stars in the LMC}
{
\hline
\noalign{\vskip3pt}
\multicolumn{1}{c}{Direction}& $\langle I\rangle$  & $\sigma^{\rm STAT}$ &
$\sigma_{\rm RC}$ & \multicolumn{1}{c}{$N_{\rm stars}$} & $E(B-V)$ & 
$\langle I_0\rangle$ & [Fe/H] \\
\noalign{\vskip3pt}
\hline
\noalign{\vskip3pt}
SL8    & 18.197 & 0.008 & 0.148 & 1345 & 0.105 & 17.991 & $-0.35$\\
SL509  & 18.039 & 0.007 & 0.127 &  784 & 0.029 & 17.982 & $-0.50$\\
SL262  & 18.037 & 0.019 & 0.138 &  223 & 0.026 & 17.986 & $-0.50$\\
SL842  & 18.001 & 0.021 & 0.130 &  157 & 0.054 & 17.895 & $-0.55$\\
SL388  & 18.053 & 0.009 & 0.114 &  749 & 0.042 & 17.971 & $-0.60$\\
SL862  & 18.143 & 0.009 & 0.150 & 1063 & 0.119 & 17.910 & $-0.60$\\
OHSC33 & 18.081 & 0.013 & 0.155 &  660 & 0.100 & 17.885 & $-0.65$\\
SL817  & 18.112 & 0.008 & 0.143 & 1427 & 0.101 & 17.914 & $-0.70$\\
IC2134 & 18.116 & 0.007 & 0.105 &  864 & 0.078 & 17.963 & $-0.70$\\
\hline}

\Subsection{LMC Red Clump Stars}

Table~1 summarizes the basic properties of field RC stars in nine fields
around star clusters distributed in different parts of the LMC halo. The
interstellar extinction in these directions is small, thus minimizing
uncertainties of extinction-free photometry. All lines-of-sight are far
enough from the LMC center so the reddening could be reliably determined
from the COBE/DIRBE maps of Schlegel, Finkbeiner and Davis (1998). The
{\it I}-band interstellar extinction was calculated using the standard
extinction curve ($A_I=1.96\cdot E(B-V)$). We assumed uncertainty of the
reddening value equal to $\pm0.02$~mag.

$\langle I\rangle$ of RC stars was derived in similar manner as in
Udalski \etal (1998). About 160--1400 field red giants were used for its
determination in our fields. Statistical uncertainty of $\langle
I\rangle$ is usually below 0.01~mag so the main contribution to the
extinction free magnitude, $\langle I_0\rangle$,   error budget  comes
from the interstellar reddening uncertainty. Dispersion of brightness of
field RC stars, $\sigma_{\rm RC}$, is typically below 0.15~mag similar
to the local RC stars.

Unfortunately metallicity of the LMC field RC stars is not known so
precisely as that of the Hipparcos local giants. Metallicity of field
giants in our nine lines-of-sight was determined by Bica \etal (1998,
herafter  BGDCPS) using Washington photometry. It ranges from $-0.7$~dex
to $-0.35$~dex (Table~1) with typical error of 0.1~dex of internal
determination, and the total error of $\approx0.2$~dex. Due to large
errors it is not clear if the observed  dispersion of metallicity in
different parts of the LMC is real or results from the uncertainty of
the method. It is also not clear whether the absolute numbers are
correct, \eg metallicity determined using the same method for several
LMC clusters by BGDCPS is on average by $\approx 0.2$~dex lower than
spectroscopic metallicities of the LMC clusters determined by Olszewski
\etal (1991). This point should be cleared up in the near future with
direct spectroscopic observations of the LMC RC stars with new 8-m class
telescopes.

\begin{figure}[htb]
\hglue 2.0cm\psfig{figure=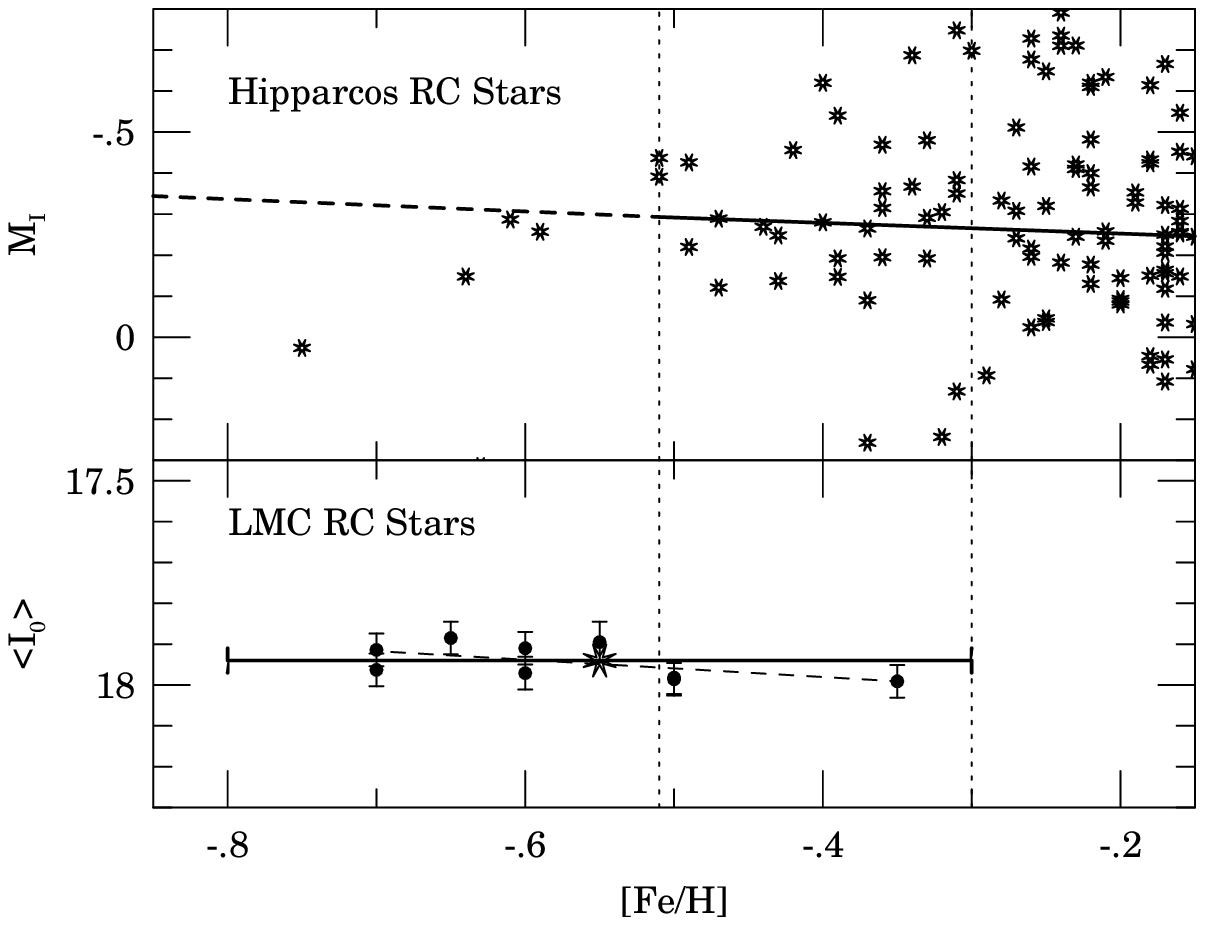,bbllx=55pt,bblly=165pt,bburx=405pt,bbury=435pt,width=8.5cm,clip=}
\vspace*{0pt}
\FigCap{
$M_I$ of Hipparcos RC giants (upper panel) and $\langle I_0\rangle$ of
RC stars in nine fields in the LMC (lower panel) plotted as a function
of metallicity.  Asterisk in the lower panel denotes the mean
metallicity of RC stars in the LMC and the horizontal bar spread of
metallicity. Dotted lines indicate the range where metallicities of the
local and LMC stars overlap. Solid thick line in the upper panel marks
the derived relation of $M_I$ on [Fe/H] (Eq.~1) while dashed line is its
continuation resulting from Girardi's (1999a) models. Dashed line in the
lower panel indicates possible relation for the LMC RC stars.
}
\end{figure}

Lower panel of Fig.~3 presents $\langle I_0 \rangle$ of RC stars in our
fields in the LMC as a function of BGDCPS [Fe/H]. Large asterisk
indicates the mean metallicity and brightness of the entire sample. 
Horizontal bar corresponds to the typical range of metallicities of
field giants, $\approx0.5$~dex, as determined by BGDCPS. It is worth
noting that if one assumes that dispersion of metallicity in the LMC
fields is real and BGDCPS determinations are correct, at least
differentially, then the observed trend of variation of $\langle I_0
\rangle$ with [Fe/H] is similar to the local sample of RC stars. The
formal fit of a straight line gives the slope equal to
$0.21\pm0.12$~mag/dex. Although one should treat this figure with
caution  it is encouraging that it is in good agreement with that
resulting from analysis of the local RC stars.

\Subsection{Distance Modulus to the LMC}

The range of metallicities among nearby RC stars is wide and covers
$-0.6<{\rm [Fe/H]}<+0.2$~dex. This is a very fortunate situation,
because this range partially overlaps at the low [Fe/H] end with 
metallicity of the RC stars in the LMC.   In the upper panel of Fig.~3
$M_I$ of Hipparcos RC stars  is plotted as a function of [Fe/H]. Solid
line marks relation of $M_I$ \vs [Fe/H] given by Eq.~(1). Two dotted
vertical lines indicate the range where metallicities of the LMC and
local Hipparcos RC stars overlap. We remind here that we have adopted
BGDCPS metallicities of the LMC RC stars, and if they are too low, what
is likely, the overlap of metallicities of both populations can be much
larger.

The mean metallicity of the LMC RC stars lies outside the overlap
region. Therefore, to derive $(m-M)_{\rm LMC}$  we have to slightly
extrapolate $M_I$ \vs [Fe/H] relation given by Eq.~(1).  $\langle
M_I\rangle$ of the Hipparcos RC giants extrapolated to the metallicity
of $-0.55$~dex is equal to $\langle M_I\rangle=-0.30\pm0.04$~mag. With
$\langle I_0 \rangle=17.94\pm0.05$~mag of RC stars in our nine
lines-of-sight in the LMC, this  immediately leads to $(m-M)_{\rm
LMC}=18.24\pm0.08~{\rm mag}$.

Can this result be severely affected by extrapolation? That could only
be possible if $M_I$ \vs [Fe/H] relation of RC stars for metallicities
in the range of $-0.9<{\rm [Fe/H]}<-0.5$, \ie not covered by Hipparcos
stars,  would behave extraordinarily. In particular it would have to be
extremely steep in this range to narrow the gap between our result and
the "long" distance modulus of $(m-M)_{\rm LMC}=18.50$. However, this is
not the case: we already mentioned that if dispersion of metallicity in
the LMC fields as measured by BGDCPS is real then the slopes of $M_I$
\vs [Fe/H] relations in the LMC and the local sample are similar. Also
theoretical modeling of RC provides similar arguments. For instance,
models of Girardi (1999a, Fig.~4) indicate that for RC stars of age
2--8~Gyr the mean slope of the $M_I$ \vs [Fe/H] relation in the range of
metallicities of $-1.0<{\rm [Fe/H]}<-0.4$ is about 0.15~mag/dex in
excellent agreement with our empirical data (theoretical relation is
plotted with dashed line in the upper panel of Fig.~3 as a continuation
of our empirical relation). Thus large uncertainty of the distance
modulus due to extrapolation of Eq.~(1) is very unlikely.

Possible differences of ages of both populations can be another
potential source of uncertainty of the derived $(m-M)_{\rm LMC}$.
Empirical study of this effect based on analysis of clusters in the
Magellanic Clouds showed that it is practically negligible for RC stars
of age within 2--10~Gyr (Udalski 1998b). On the other hand  analysis of
a few Galactic open clusters by Sarajedini (1999) suggests that RC stars
older than $\approx 5$~Gyr become fainter. Without going into detailed
discussion of these results we only note here that analysis of Galactic
clusters requires very precise distance determination what is difficult
and that two old clusters claimed to have fainter RC (Be39, NGC188) have
actually very sparse population of RC stars consisting of only a few
stars. However, despite differences which deserve further studies, both
Udalski (1998b) and Sarajedini (1999) data show that for the age range
of 2--5~Gyr $\langle M_I\rangle$ of RC giants is constant within
$\pm0.05$~mag. The age of the LMC RC stars is within this range (BGDCPS)
and the vast majority of the local RC stars are younger than 4~Gyr
(Girardi 1999b). To be on the safe side we included  to the final error
budget uncertainty of $\pm0.05$~mag for possible differences of age
between both populations. We may conclude that the derived distance
modulus to the LMC is largely free from population uncertainties. Also
small interstellar extinction assures that the result is sound. Had
there been any additional LMC extinction in the LMC fields, on top of
that given by Schlegel \etal (1998), $(m-M)_{\rm LMC}$ would be reduced
to even smaller value.

\Acknow{We would like to thank the anonymous referee whose critical
remarks  allowed us to significantly improve the manuscript. We are very
grateful to Mr.\ K.~\.Zebru\'n for collecting observations of the LMC
fields. We thank Drs.\ B.\ Paczy\'nski, K.Z.\ Stanek, M.\ Kubiak and M.\
Szyma\'nski for many discussions and important suggestions. The paper
was partly supported by the grants: Polish KBN 2P03D00814 and NSF
AST-9820314.}


\begin{references}
\refitem{Bica, E., Geisler, D., Dottori, H., Clari\'a, J.J., Piatti, A.E.,
and Santos Jr, J.F.C.}{1998}{\AJ}{116}{723 (BGDCPS)}
\refitem{Castellani, V., Degl'Innocenti, S., Girardi, L., Marconi,
M., Prada Moroni, P.G., and Weiss, A.}{1999}{\AA}{~}{in press,
astro-ph/9911432}
\refitem{Dominguez, I., Chieffi, A., Limongi, M., and Straniero, O.}
{1999}{\ApJ}{524}{226}
\refitem{Girardi, L., Groenewegen, M.A.T, Weiss, A., and Salaris,
M.}{1998}{\MNRAS}{301}{149}
\refitem{Girardi, L.}{1999a}{\MNRAS}{308}{818}
\refitem{Girardi, L.}{1999b}{~}{~}{astro-ph/9912309}
\refitem{Horner, D. \etal}{1999}{~}{~}{astro-ph/9907213}
\refitem{Landolt, A.U.}{1992}{\AJ}{104}{372}
\refitem{McWilliam, A.}{1990}{\ApJS}{74}{1075}
\refitem{McWilliam, A.}{1997}{AAR\&A}{35}{503}
\refitem{Olszewski, E.W., Schommer, R.A., Suntzeff, N.B., and Harris,
H.C.}{1991}{\AJ}{101}{515}
\refitem{Paczy\'nski B., and Stanek, K.Z.}{1998}{\ApJL}{494}{L219 (PS)}
\refitem{Paczy\'nski B., Udalski, A., Szyma\'nski, M., Kubiak, M.,
Pietrzy\'nski, G.,  Soszy\'nski, I., Wo\'zniak, P., and {\.Z}ebru\'n,
K.}{1999} {\Acta}{49}{319}
\refitem{Perryman, M.A.C. \etal}{1997}{\AA}{323}{L49}
\refitem{Sarajedini, A.}{1999}{\AJ}{118}{2321}
\refitem{Schlegel, D.J., Finkbeiner, D.P., and Davis, M.}{1998}{\ApJ}{500}{525}
\refitem{Stanek, K.Z, Zaritsky, D., and Harris, J.}{1998}{\ApJL}{500}{L141}
\refitem{Taylor, B.J.}{1999}{\AAS}{135}{75}
\refitem{Udalski, A., Kubiak, M., and Szyma\'nski, M.}{1997}{\Acta}{47}{319}
\refitem{Udalski, A.}{1998a}{\Acta}{48}{113}
\refitem{Udalski, A.}{1998b}{\Acta}{48}{383}
\refitem{Udalski, A., Szyma\'nski, M., Kubiak, M., Pietrzy\'nski, G.,
Wo\'zniak, P., and {\.Z}ebru\'n, K.}{1998}{\Acta}{48}{1}

\end{references}
\end{document}